\documentclass[12pt]{article}
\usepackage{amssymb}
\usepackage{amsmath}
\usepackage{color}

\usepackage{hyperref}
\usepackage{bbm}
\usepackage{authblk}

\newcommand{\tr}{\mathrm{Tr}}

\title{Spin-density and Vorticity Contribution to the Cosmological Background}

\author[1]{Dario Capasso\footnote{dario.capass@gmail.com}}
\affil[1]{Physics Department, City College of New York - CUNY, New York, NY 10031}
\begin{document}

\maketitle
%\hfill\raisebox{300pt}[0pt][0pt]{CCNY-HEP-12/10}
\begin{abstract}
Relativistic non-Abelian spinning fluids can be formulated in group theory language, where the corresponding Mathisson-Papapetrou equation for spinning fluids can be obtained in terms of a specific de Sitter group contraction. This framework is very suitable for studying the effects of a spinning fluid of matter with vorticity and a spin density in a cosmological background.
% In particular the contribution to the cosmological background from a stationary spin current is studied.
\end{abstract}

\tableofcontents

%%%%%%%%%%%%%%%%%%%%%%%%%%%%%%%%%%%%
\section{Introduction}
\label{sec:intro}
%%%%%%%%%%%%%%%%%%%%%%%%%%%%%%%%%%%%\\

On the largest scale, the universe is described by the homogeneous and isotropic Friedman-Robertson-Walker (FRW) metric; this approximate model does not capture the complexity and richness of our universe. The spectrum of the anisotropies has been mapped several times and, recently, by the ESA's Plank satellite, which produced the picture of Cosmic Microwave Background (CMB) with the highest precision available until now. Our universe is indeed the consequence of the presence of anisotropies in the early structure of the universe.

The universe can then be described on a large scale through a perturbed FRW metric, on one side, and through inhomogeneities and anisotropies of a fluid permeating it. In particular we are interested in relating the vorticity and the spin-density of a relativistic matter fluid with spin to the perturbed metric.

We shall consider the background metric to be flat
\begin{equation}\label{eqn:flatmetric}
ds^{2}
=\bar{g}_{\alpha\beta}dx^{\alpha}dx^{\beta}
=a^{2}(\tau)\eta_{\mu\nu}dx^{\mu}dx^{\nu},
\end{equation}
which is also confirmed by recent observations \cite{Ade:2013zuv}. Then small inhomogenous perturbations on top of the metric $\bar{g}_{\alpha\beta}$ will be considered to describe the anisotropy of the universe
\begin{equation}\label{eqn:perturbedmetric}
ds^{2}
=[\bar{g}_{\alpha\beta}+\delta g_{\alpha\beta}]dx^{\alpha}dx^{\beta}
=a^{2}(\tau)[(1+2A)d\tau^{2}-2B_{i}dx^{i}d\tau-(\delta_{ij}+h_{ij})dx^{i}dx^{j}].
\end{equation}
The first order perturbations $A$, $B_{i}$, and $h_{ij}$ can be better characterized performing a scalar-vector-tensor (SVT) decomposition: we have a scalar $A$, a spatial vector $B_{i}$ that can be decomposed into a gradiant of a scalar plus a divergenceless vector $\hat{B}_{i}$
\[
B_{i}=\partial_{i}\underbrace{B}_{scalar}+\underbrace{\hat{B}_{i}}_{vector} \qquad
\textrm{with }
\partial_{i}\hat{B}_{i}=0,
\]
and a symmetric spatial tensor $h_{ij}$ that can be decomposed in 2 scalars, $C$ and $E$ , a divergenceless spatial vector $\hat{E}_{i}$ and a divergenceless traceless symmetric tensor $\hat{E}_{ij}$
\[
h_{\underline{ij}}
=\underbrace{
2C\delta_{ij}
+2\partial_{\langle i}\partial_{j\rangle}E
}_{scalars}
+\underbrace{2\partial_{(i}\hat{E}_{j)}}_{vector}
+\underbrace{2\hat{E}_{ij}}_{tensor}
\]\[
\textrm{with}\quad
\delta^{ij}\hat{E}_{ij}=0,\qquad
\delta^{ij}\partial_{i}\hat{E}_{j}=0 \qquad
\textrm{and}\quad
\delta^{ki}\partial_{k}\hat{E}_{ij}=0
\]
where
\[
\partial_{\langle\underline{i}}\partial_{\underline{j}\rangle}
\equiv\left(
\partial_{\underline{i}}\partial_{\underline{j}}
-\frac{1}{3}\delta_{\underline{ij}}\nabla^{2}
\right).
\]
The 10 degrees of freedom of the metric have thus been decomposed into 4 + 4 + 2 SVT degrees of freedom.
\begin{itemize}
\item scalars: $A$, $B$, $C$, $E$;
\item vectors: $\hat{B}_{i}$, $\hat{E}_{i}$;
\item tensors: $\hat{E}_{ij}$.
\end{itemize}
Under a small (comparable to the order of the perturbations) transformation of coordinates $x^{\mu}\to x^{\mu}+(T,\partial^{i}L+\hat{L}^{i}$ (with $\partial_{i}\hat{L}_{i}=0$ to decompose the spatial transformation in a scalar plus a divergenceless vector) the above fields transform as
\begin{eqnarray}
A&\to& A-\dot{T}-HT\label{eqn:Atrans}\\
B&\to& B+T-\dot{L} \qquad
\hat{B}_{i}\to\hat{B}_{i}-\dot{\hat{L}}_{i}\label{eqn:B_itrans}\\
C&\to& C-HT-\frac{1}{3}\nabla^{2}L\\
E&\to& E-L \qquad
\hat{E}_{i}\to\hat{E}_{i}-\hat{L}_{i} \qquad
\hat{E}_{ij}\to\hat{E}_{ij}
\end{eqnarray}
Therefore there is ambiguity in defining the above fields. Such an approach then introduces gauge issues. The spurious degrees of freedom can be removed by a choice of gauge as we shall do in section~\ref{sec:vorticity}. On the other hand, a different approach that will be partially used in appendix~\ref{app:curvature} is to use the Bardeen's variables, written below, which represent an invariant set of variables describing the physical 2 + 2 + 2 SVT degrees of freedom:
\begin{eqnarray}
\Phi
&\equiv& -C-H(B-\dot{E})+\frac{1}{3}\nabla^{2}E, \qquad
\hat{\Phi}_{i}
\equiv \dot{\hat{E}}_{i}-\hat{B}_{i}\nonumber\\
\Psi
&\equiv& A+H(B-\dot{E})+(\dot{B}-\ddot{E}), \qquad
\hat{E}_{ij}\label{eqn:Bardeen}
\end{eqnarray}
where $\dot{\phantom{}}$ is the partial derivative with respect to the cosmological time $\tau$, and $H=\dot{a}/a$ is the Hubble constant.

To parametrize matter we shall use a group theoretical model to describe fluids recently introduced in \cite{Capasso:2012dr}.   This group theoretical approach was developed in \cite{Bistrovic:2002jx,Jackiw:2004nm} and relies on symmetry principles which make almost straightforward to write the action for a fluid.

For a general gauge group $G$, describing the local symmetries of a fluid, the action describing the fluid is given by 
\[
S=-i\int\sum_{(i)}j_{(i)}^\mu\mbox{Tr}(K_{(i)}g^{-1}D_\mu g)-\int F(n_1,n_2,\dots)+S_{YM}(A)
\]
where the $K_{(i)}$ are diagonal generators of $G$ and $j_{(i)}^{\mu} j_{(i)\mu}=n_{(i)}^{2}$ with $i$ indexing the rank of the group $G$. This technique as recently been studied and discussed in several topics in \cite{Nair:2011mk,Karabali:2014vla,Capasso:2013jva}.

Following \cite{Capasso:2012dr}, we can describe a fluid with spin choosing the group $G$ to be deSitter. The Poincar\'e group is an alternate choice, but produces overconstrained models (see \cite{Capasso:2012dr} for more details). The action describing the fluids is
\begin{equation}\label{eqn:dSaction}
S=
-\int\det e\left[
ij_{(m)}^{\mu}\tr(\alpha_{(m)}T_{0}g^{-1}\nabla_{\mu}g)
+ij_{(s)}^{\mu}\tr(\alpha_{(s)}T_{23}g^{-1}\nabla_{\mu}g)
+F(n_{(m)},n_{(s)})
\right]
\end{equation}
where $g$ is a $SO(4,1)$-valued field and $\alpha_{(m)}$ and $\alpha_{(s)}$ are constants respectively for the mass and spin coupling. The choice of the diagonal Lie algebra elements of $\mathbf{so}(4,1)$ is taken considering that we are interested in studying matter fluids ($T_{0}$ generates a time-like velocity) with spin, generated by $T_{23}$ (see appendix~\ref{app:dSgenerators} for a brief description of the generators of $\mathbf{so}(4,1)$).

Action (\ref{eqn:dSaction}) is invariant separately under $g\to ge^{-i\gamma_{0}T^{0}}$ and $g\to ge^{-i\gamma_{23}T^{23}}$ with $\gamma_{0}$ and $\gamma_{23}$ constants; the associated conserved currents are $j_{(i)}^{\mu}$ with
\begin{equation}\label{eqn:dSDj=0}
\nabla_{\mu}j_{(i)}^{\mu}=0
\end{equation}
where $i=\{m,s\}$. The explicit form of the currents, obtained by varying action (\ref{eqn:dSaction}) with respect to the current $j_{(i)\mu}$ itself, is
\begin{equation}\label{eqn:dSj_i}
j_{(i)\mu}=-i\frac{n_{(i)}}{F_{(i)}}\tr[K_{(i)}g^{-1}\nabla_{\mu}g]
\end{equation}
with $F_{(i)}\equiv \partial F/\partial n_{(i)}$, $K_{(m)}=\alpha_{(m)}T_{0}$ and $K_{(s)}=\alpha_{(s)}T_{23}$. The remaining equation of motion is obtained by varying the action with respect to field $g$
\begin{equation}\label{eqn:dSDJ=0}
\nabla_{\mu}\left(
j_{(m)}^{\mu}gK_{(m)}g^{-1}
+j_{(s)}^{\mu}gK_{(s)}g^{-1}
\right)=
0
\end{equation}
that states the conservation of the $\mathbf{so}(4,1)$-current
\begin{equation}\label{eqn:dSJ}
J^{\mu}
=j_{(m)}^{\mu}gK_{(m)}g^{-1}
+j_{(s)}^{\mu}gK_{(s)}g^{-1}.
\end{equation}

In section~\ref{sec:Spinning_fluids_FRW}, the expressions for the perturbed currents are derived. In section~\ref{sec:EoM} we derive the set of equations which describe the fluid and the perturbed metric. In appendix~\ref{app:curvature} the unperturbed Christoffel symbols, Riemannian tensor and the Ricci scalar for the metric (\ref{eqn:flatmetric}) are derived, as well as their first order corrections from the perturbed metric (\ref{eqn:perturbedmetric}). Notice that no gauge choice as been made in appendix~\ref{app:curvature}. In appendix~\ref{app:dSgenerators} we list the convention used in this article for the representation of the Lie algebra associated with the de Sitter group.

From now on, to easily distinguish between group and tensor indexes, we shall use Greek letters as tensor indexes, \underline{underlined} Latin letters for spatial tensor indexes, and Latin letters for group indexes. In general, given an expression, we will use the prefix $\delta$ to express that is the first-order component and we shall use a $\bar{\phantom{a}}$ to indicate that it is its zero-order part.

%%%%%%%%%%%%%%%%%%%%%%%%%%%%%%%%%%%%
\section{Spinning Fluids in FRW background}
\label{sec:Spinning_fluids_FRW}
%%%%%%%%%%%%%%%%%%%%%%%%%%%%%%%%%%%%

To reproduce the correct dynamics for the fluid the total momentum and the total spin should be conserved. For this to happen the theory should reproduce the right Casimirs. The $SO(4,1)$ Casimirs as well as the commutation relations between the Lie generators of the group are different from the one for the Poincar\'e group, which is what we need. Therefore, to reproduce the right dynamic we need to consider a group contraction of the deSitter group to the Poincar\'e group. In \cite{Capasso:2012dr} it is shown that the right group contraction procedure is given by considering group elements of the form
\[
g=\Lambda e^{-i\frac{\theta^{a}T_{a}}{R}}
\]
with $R$ the deSitter radius. By considering the contraction $R\to\infty$ we reproduce the Poincar\'e commutation relations were the $T_{a}$ will act as the momentum generators.
The above procedure will not ba able to reproduce the needed Casimirs; for this to happen the above spin coupling constant $\alpha_{(s)}$ should be rescaled to $\alpha_{(s)}/R$. This must be done at the level of the action. Therefore the expressions for the currents that we will consider from now are
\begin{eqnarray}
j_{(m)\mu}
&=&-i\frac{n_{(m)}\alpha_{(m)}}{F_{(m)}}\tr[T_{0}g^{-1}\nabla_{\mu}g]\\
j_{(s)\mu}
&=&-i\frac{n_{(s)}\alpha_{(s)}}{F_{(s)}R}\tr[T_{23}g^{-1}\nabla_{\mu}g].
\end{eqnarray}
Under the group contraction the currents reduce to
\begin{eqnarray}
j_{(m)\mu}
%=-\frac{n_{(m)}\alpha_{(m)}}{F_{(m)}R}\tr[T_{0}\partial_{\mu}\theta^{a}T_{a}]
%-i\frac{n_{(m)}\alpha_{(m)}}{F_{(m)}}\tr[hT_{0}h^{-1}\Lambda^{-1}\partial_{\mu}\Lambda]
%-i\frac{n_{(m)}\alpha_{(m)}}{F_{(m)}}\tr[hT_{0}h^{-1}\Lambda^{-1}w_{\mu}\Lambda]
%\]\[
%=\frac{n_{(m)}\alpha_{(m)}}{4F_{(m)}R}\partial_{\mu}\theta_{0}
%-\frac{n_{(m)}\alpha_{(m)}}{F_{(m)}R}\tr[\theta^{a}[T_{a},T_{0}]\Lambda^{-1}\partial_{\mu}\Lambda]
%-\frac{n_{(m)}\alpha_{(m)}}{F_{(m)}R}\tr[\theta^{a}[T_{a},T_{0}]\Lambda^{-1}w_{\mu}\Lambda]
%\]\[
%=\frac{n_{(m)}\alpha_{(m)}}{4F_{(m)}R}\partial_{\mu}\theta_{0}
%-i\frac{n_{(m)}\alpha_{(m)}}{2F_{(m)}R}\tr[\theta^{a}T_{a0}\Lambda^{-1}\partial_{\mu}\Lambda]
%-i\frac{n_{(m)}\alpha_{(m)}}{2F_{(m)}R}\tr[\theta^{a}T_{a0}\Lambda^{-1}w_{\mu}\Lambda]
%\]\[
&=&\frac{n_{(m)}\alpha_{(m)}}{4F_{(m)}R}\partial_{\mu}\theta_{0}
-\frac{n_{(m)}\alpha_{(m)}}{4F_{(m)}R}\theta^{a}(\Lambda^{-1}\nabla_{\mu}\Lambda)_{a0}
+\mathcal{O}(R^{-3})\\
j_{(s)\mu}
%=-i\frac{n_{(s)}\alpha_{(s)}}{F_{(s)}R}\tr[T_{23}\Lambda^{-1}\nabla_{\mu}\Lambda]
%=-\frac{n_{(s)}\alpha_{(s)}}{2F_{(s)}R}\Lambda_{a2}\nabla_{\mu}\Lambda^{a}_{\phantom{-}3}
%=-\frac{n_{(s)}\alpha_{(s)}}{2F_{(s)}R}\Lambda_{a2}[
%\partial_{\mu}\Lambda^{a}_{\phantom{-}3}
%+w^{a}_{\mu b}\Lambda^{b}_{\phantom{-}3}
%]
&=&-\frac{n_{(s)}\alpha_{(s)}}{2F_{(s)}R}\Lambda_{a2}
\partial_{\mu}\Lambda^{a}_{\phantom{-}3}
+\mathcal{O}(R^{-3})
\end{eqnarray}

Going back to the case in study, in \cite{Capasso:2012dr} it has been shown how our group theoretical model describing fluids with spins reproduces the expected trivial solution describing an homogeneous and isotropic fluid of matter with  no vorticity and no spin density in the case of a purely FRW background. That example will be expanded in this and the following sections to take vorticity and spin-density into account as first order perturbation.

The group-valued field describing the trivial case of a matter fluid permeating an FRW spacetime, as described in \cite{Capasso:2012dr}, takes the following expression
\[
g=\Lambda^{}_{FRW}h^{}_{FRW}
\]
with
\[
(\Lambda^{}_{FRW})_{ab}=\left(\begin{array}{cc}
1 & 0\\
0 & \mathcal{A}
\end{array}\right) \qquad\textrm{with}\quad \mathcal{A}\in SO(3)
\]
and
\[
h^{}_{FRW}=e^{-i\frac{\theta^{0}}{R}T_{0}}.
\]
The matrix $\mathcal{A}$ is a generic $SO(3)$-valued expression because no spin ($\alpha_{(s)}=0$) and no vorticity ($\theta^{1}=\theta^{2}=\theta^{3}=0$) were considered. In this article we want to derive equations for the matrix $\mathcal{A}$ considering $\alpha_{(s)}$, $\theta^{1}$, $\theta^{2}$, and $\theta^{3}$ as perturbation parameters of the same order.

%%%%%%%%%%%%%%%%%%%%%%%%%%%%%%%%%%%%
\subsection{Vorticity}
\label{sec:vorticity}
%%%%%%%%%%%%%%%%%%%%%%%%%%%%%%%%%%%%

In this section we will describe, using a perturbative approach, how the mass transport current is modified considering the presence of vorticity ($\delta\theta^{1},\;\delta\theta^{2},\;\delta\theta^{3}\neq0$) treated as a first order correction:
\begin{eqnarray}
j_{(m)\mu}
&=&\frac{\bar{n}_{(m)}\alpha_{(m)}}{4F_{(m)}R}\left(1+\frac{\delta n_{(m)}}{\bar{n}_{(m)}}\right)\left[
\partial_{\mu}(\theta_{0}+\delta\theta_{0})
-\delta\theta^{a}(\Lambda^{-1}\nabla_{\mu}\Lambda)_{a0}
\right]\nonumber\\
%\]\[
%=\frac{\bar{n}_{(m)}\alpha_{(m)}}{4F_{(m)}R}\left(1+\frac{\delta n_{(m)}}{\bar{n}_{(m)}}\right)\left[
%\partial_{\mu}(\theta_{0}+\delta\theta_{0})
%-\delta\theta^{i}\Lambda_{ji}e^{j}_{\alpha}(\partial_{\mu}e^{\alpha}_{0}+\Gamma^{\alpha}_{\mu\beta}e^{\beta}_{0})
%\right]
%\]\[
&=&\frac{\bar{n}_{(m)}\alpha_{(m)}}{4F_{(m)}R}\left(1+\frac{\delta n_{(m)}}{\bar{n}_{(m)}}\right)\left[
\partial_{\mu}(\theta_{0}+\delta\theta_{0})
-\frac{\dot{a}}{a^{2}}e^{j}_{\mu}\Lambda_{ji}\delta\theta^{i}
\right]\label{eqn:j_perturbed}
\end{eqnarray}
where $\delta\theta_{0}$ takes into account the cosmological-time perturbation, $\delta n_{(m)}$ the fluid density\marginpar{check} perturbation, and $\delta\theta^{i}$ are the vorticity parameters.

From the normalization of $u_{(m)}^{\mu}$
\[
0
={\bar{u}_{(m)}}^{\mu}{\bar{u}_{(m)}}^{\nu}\delta g_{\mu\nu}
+2\bar{u}_{(m)\mu}\delta{u_{(m)}}^{\mu}
=({\bar{u}_{(m)}}^{\tau})^{2}\delta g_{\tau\tau}
+2\bar{u}_{(m)\tau}\delta{u_{(m)}}^{\tau}
\Rightarrow
\]\[
\Rightarrow
\delta{u_{(m)}}^{\tau}
=-\frac{{\bar{u}_{(m)}}^{\tau}}{2a^{2}}\delta g_{\tau\tau}
=-\frac{A}{a}
\]
we obtain an expression for the four-velocity $\delta u_{(m)}^{\mu}$
\[
{u_{(m)}}^{\mu}
=a^{-1}[1-A,v^{i}] \qquad
{u_{(m)}}_{\mu}
=a[1+A,-(v_{i}+B_{i})]
\]
where $v_{\underline{i}}=v^{\underline{j}}\delta_{\underline{ji}}$.

Comparing the expression above for $u_{(m)}^{\mu}$ and the expression~(\ref{eqn:j_perturbed}) we find
\[
\partial_{\tau}\theta_{0}
=\frac{4F_{(m)}R}{\alpha_{(m)}}a \qquad
\partial_{\underline{j}}\theta_{0}
=0
\qquad\Rightarrow\qquad
\theta_{0}
=\int d\tau\frac{4F_{(m)}R}{\alpha_{(m)}}a \qquad
\]\[
\partial_{\tau}\delta\theta_{0}
=\frac{4F_{(m)}R}{\alpha_{(m)}}aA \qquad
\partial_{\underline{j}}\delta\theta_{0}
-\frac{\dot{a}}{a^{2}}e^{j}_{\underline{j}}\Lambda_{ji}\delta\theta^{i}
=-\frac{4F_{(m)}R}{\alpha_{(m)}}a(v_{\underline{j}}+B_{\underline{j}})
\]

At this point we can start making a gauge choice. To simplify the expression for the linearized Einstein equations (see appendix~\ref{app:curvature}) we choose the Newtonian gauge ($B=E=0$ and $\hat{B}_{\underline{i}}=0$). The four-velocity then takes the following form
\begin{equation}
u_{(m)\tau}
=\frac{\alpha_{(m)}}{4F_{(m)}R}\partial_{\tau}\theta_{0} \qquad
u_{(m)\underline{j}}
=\frac{\alpha_{(m)}}{4F_{(m)}R}\left[
\partial_{\underline{j}}\theta_{0}
-\frac{H}{a}e^{j}_{\underline{j}}\Lambda_{ji}\delta\theta^{i}
\right]
\end{equation}
giving
\begin{equation}\label{eqn:v}
v_{\underline{j}}
=\frac{\alpha_{(m)}}{4F_{(m)}R}\frac{1}{a}\left[
\frac{H}{a}e^{j}_{\underline{j}}\Lambda_{ji}\delta\theta^{i}
-\partial_{\underline{j}}\theta_{0}
\right], \qquad
v^{\underline{j}}
=\frac{\alpha_{(m)}}{4F_{(m)}R}\left[
He_{j}^{\underline{j}}\Lambda^{j}_{\phantom{-}i}\delta\theta^{i}
-\frac{\delta^{\underline{ji}}}{a}\partial_{\underline{i}}\theta_{0}
\right]
\end{equation}

The vorticity of the mass current is given by
\begin{equation}
\omega_{\nu\mu}
=\nabla_{[\nu}\left(\frac{F_{(m)}}{n_{(m)}}j_{(m)\mu]}\right)
=-\frac{\alpha_{(m)}}{4R}\partial_{[\nu}^{\phantom{-}}\left[
\frac{\dot{a}}{a^{2}}e^{j}_{\mu]}\Lambda_{ji}\delta\theta^{i}
\right]
\end{equation}

\[
\omega_{\underline{j}\tau}
=\frac{\alpha_{(m)}}{8R}\partial_{\tau}^{\phantom{-}}\left[
\frac{\dot{a}}{a}\delta^{j}_{\underline{j}}\Lambda_{ji}\delta\theta^{i}
\right]
\qquad
\omega_{\underline{ij}}
=-\frac{\alpha_{(m)}}{4R}\partial_{[\underline{i}}^{\phantom{-}}\left[
\frac{\dot{a}}{a}\delta^{j}_{\underline{j]}}\Lambda_{jk}\delta\theta^{k}
\right]
\]

The above expression can be rewritten in terms of the charges associated with the mass current
\[
Q^{ab}_{(m)}
=-i\alpha_{(m)}\tr[gT_{0}g^{-1}T^{ab}]
=-i\alpha_{(m)}\tr[\Lambda hT_{0}h^{-1}\Lambda^{-1}T^{ab}]
%\]\[
%=-i\alpha_{(m)}\tr[\Lambda T_{0}\Lambda^{-1}T^{ab}]
%-\frac{\alpha_{(m)}\theta^{e}}{R}\tr[\Lambda[T_{e},T_{0}]\Lambda^{-1}T^{ab}]
%\]\[
%=-\frac{\alpha_{(m)}\theta^{e}}{2R}\tr[\Lambda T_{e0}\Lambda^{-1}T^{ab}]
%=-\frac{\alpha_{(m)}\theta^{e}}{2R}\Lambda^{c}_{\phantom{-}e}\Lambda^{d}_{\phantom{-}0}\tr[T^{ab}T_{cd}]
%=\frac{\alpha_{(m)}\theta^{e}}{4R}\Lambda^{[a}_{\phantom{-}e}\Lambda^{b]}_{\phantom{-}0}
%\]\[
=\frac{\alpha_{(m)}\delta\theta^{e}}{8R}[
\Lambda^{a}_{\phantom{-}e}\delta^{b}_{0}
-\Lambda^{b}_{\phantom{-}e}\delta^{a}_{0}
]
\]\[
Q^{i0}_{(m)}=-Q^{0i}_{(m)}=\frac{\alpha_{(m)}}{8R}\Lambda^{i}_{\phantom{-}j}\delta\theta^{j} \qquad
Q^{ij}_{(m)}=0
\]
In particular we have
\[
v^{\underline{j}}
=2aHQ_{(m)}^{\underline{j}\tau}
-\frac{\alpha_{(m)}}{4F_{(m)}R}\frac{\delta^{\underline{ji}}}{a}\partial_{\underline{i}}\theta_{0}
\]

%%%%%%%%%%%%%%%%%%%%%%%%%%%%%%%%%%%%
\subsection{Spin-density}
\label{sec:Spindensity}
%%%%%%%%%%%%%%%%%%%%%%%%%%%%%%%%%%%%

We assume here that the spin current
\[
j_{(s)\mu}
=-i\frac{n_{(s)}\alpha_{(s)}}{F_{(s)}R}\tr[T_{23}\Lambda^{-1}\nabla_{\mu}\Lambda]
=-\frac{n_{(s)}\alpha_{(s)}}{2F_{(s)}R}\Lambda_{a2}\nabla_{\mu}\Lambda^{a}_{\phantom{-}3}
\]\[
=-\frac{n_{(s)}\alpha_{(s)}}{2F_{(s)}R}\Lambda_{a2}[
\partial_{\mu}\Lambda^{a}_{\phantom{-}3}
+w^{a}_{\mu b}\Lambda^{b}_{\phantom{-}3}
]
=-\frac{n_{(s)}\alpha_{(s)}}{2F_{(s)}R}\Lambda_{a2}
\partial_{\mu}\Lambda^{a}_{\phantom{-}3}
\]
is of the same order of the mass current $j_{(m)}^{\mu}$ or, in other words that $\bar{n}_{(s)}$ is comparable with $\bar{n}_{(s)}$. This assumption is not in contrast with the observation because the interactions are still determined by the coupling constant $\alpha_{(s)}$. As consequence we must require
\[
F_{(s)}\simeq\alpha_{(s)}.
\]

Given the generic expression for an $SO(3)$ matrix, we can write
%$\alpha$ around the x-axis, $\beta$ around the y-axis, and $\gamma$ around the z-axis.
\[
(\Lambda^{}_{FRW})_{ab}=\left(\begin{array}{cccc}
1&0&0&0\\
0& \cos\beta\cos\gamma & \cos\gamma\sin\alpha\sin\beta-\cos\alpha\sin\gamma & \cos\alpha\cos\gamma\sin\beta+\sin\alpha\sin\gamma\\
0& \cos\beta\sin\gamma & \cos\alpha\cos\gamma+\sin\alpha\sin\beta\sin\gamma & -\cos\gamma\sin\alpha+\cos\alpha\sin\beta\sin\gamma \\
0& -\sin\beta & \cos\beta\sin\alpha & \cos\alpha\cos\beta
\end{array}\right)
\]
which gives
\[
(\Lambda^{-1}\nabla_{\mu}\Lambda)_{23}
=\Lambda_{a2}\nabla_{\mu}\Lambda^{a}_{\phantom{-}3}
=\Lambda_{a2}\partial_{\mu}\Lambda^{a}_{\phantom{-}3}
=\partial_{\mu}\alpha
-\sin\beta\partial_{\mu}\gamma.
\]
The spin current can then be rewritten in terms of the $\alpha$, $\beta$, and $\gamma$ fields
\begin{equation}
j_{(s)\mu}
=-\frac{n_{(s)}\alpha_{(s)}}{2F_{(s)}R}\left(
\partial_{\mu}\alpha
-\sin\beta\partial_{\mu}\gamma
\right)
\end{equation}

The spin density can also be explicitly written in terms of the $\alpha$, $\beta$, and $\gamma$ fields
\[
Q_{(s)}^{ab}
=\frac{\alpha_{(s)}}{R}\Lambda^{a}_{\phantom{-}[2}\Lambda^{b}_{\phantom{-}3]}=
\]\[
%=\frac{\alpha_{(s)}}{2R}\left(\begin{array}{cccc}
%0&0&0&0\\
%0&0& \sin\alpha\cos^{2}\beta\sin\gamma+\cos\alpha\sin\beta\cos\gamma+\sin\alpha\sin^{2}\beta\sin\gamma & \cos\alpha\cos^{2}\beta\sin\gamma-\sin\alpha\sin\beta\cos\gamma+\cos\alpha\sin^{2}\beta\sin\gamma\\
%0& -\cos\alpha\sin\beta\cos\gamma-\sin\alpha\sin^{2}\beta\sin\gamma-\sin\alpha\cos^{2}\beta\sin\gamma &0& \cos^{2}\alpha\cos\beta\cos\gamma+\frac{1}{4}\sin2\alpha\sin2\beta\sin\gamma
%+\sin^{2}\alpha\cos\beta\cos\gamma-\frac{1}{4}\sin2\alpha\sin2\beta\sin\gamma\\
%0& \sin\alpha\sin\beta\cos\gamma-\cos\alpha\sin^{2}\beta\sin\gamma-\cos\alpha\cos^{2}\beta\sin\gamma & -\sin^{2}\alpha\cos\beta\cos\gamma+\frac{1}{4}\sin2\alpha\sin2\beta\sin\gamma
%-\cos^{2}\alpha\cos\beta\cos\gamma-\frac{1}{4}\sin2\alpha\sin2\beta\sin\gamma &0
%\end{array}\right)
%\]\[
=\frac{\alpha_{(s)}}{2R}\left(\begin{array}{cccc}
0&0&0&0\\
0&0& \sin\alpha\sin\gamma+\cos\alpha\sin\beta\cos\gamma & \cos\alpha\sin\gamma-\sin\alpha\sin\beta\cos\gamma\\
0& -\cos\alpha\sin\beta\cos\gamma-\sin\alpha\sin\gamma &0& \cos\beta\cos\gamma\\
0& \sin\alpha\sin\beta\cos\gamma-\cos\alpha\sin\gamma & -\cos\beta\cos\gamma &0
\end{array}\right)
\]
Notice that, as expected, the spin-density is contributing only to the first order and not to the zero-order.

\section{Equations of Motion}
\label{sec:EoM}
%%%%%%%%%%%%%%%%%%%%%%%%%%%%%%%%%%%%%%%%%%%

The unperturbed FRW case has been described in \cite{Capasso:2012dr} as example. The homogeneous and isotropic spacetime was trivially compatible with dust with zero vorticity and zero spin-density permeating it. In this framework a fluid  with zero vorticity and zero spin-density corresponds to a perfect fluid with
\[
\rho=F \qquad
p=F_{(m)}n_{(m)}+F_{(s)}n_{(s)}-F.
\]
For simplicity we are only going to consider a non-interacting fluid, that is dust with spin. To generalize the dust condition for a simple matter fluid without spin, $F=n$, we shall start by imposing the pressureless condition
\[
p=F_{(m)}n_{(m)}+F_{(s)}n_{(s)}-F=0
\]
with the request that at the zero order $\bar{F}(\bar{n}_{(m)},\bar{n}_{(s)})=\bar{n}_{(m)}$. Considering $\bar{n}_{(m)}$ and $n_{(s)}$ to be of the same order, the only correction comes from $\delta n_{(m)}$ and $\alpha_{(s)}$. The expression we find then is
\begin{equation}
F(n_{(m)},n_{(s)})
=n_{(m)}+\alpha_{(s)}q n_{(s)}
=\bar{n}_{(m)}+\delta n_{(m)}+\alpha_{(s)}q\bar{n}_{(s)}
\end{equation}
where $q$ is a generic constant. The above expression yields
\[
F_{(m)}=1 \qquad
F_{(s)}=\frac{\partial F}{\partial\bar{n}_{(s)}}=\alpha_{(s)}q.
\]

The first set of equations is given by the conservation of all the currents.

The mass current conservation gives
\[
0=\nabla_{\mu}j_{(m)}^{\mu}
=\bar{\nabla}_{\mu}\bar{j}_{(m)}^{\mu}
+\bar{\nabla}_{\mu}(\bar{n}_{(m)}\delta u_{(m)}^{\mu})
+\bar{\nabla}_{\mu}(\delta n_{(m)}\bar{u}_{(m)}^{\mu})
+\delta\Gamma_{\mu\alpha}^{\mu}\bar{j}_{(m)}^{\mu}
\]
From the  zero-order ($\bar{\nabla}_{\mu}\bar{j}_{(m)}^{\mu}=0$) condition one derives the expected condition
\begin{equation}
\bar{n}_{(m)}=\beta_{(m)} a^{-3}
\end{equation}
with $\beta_{(m)}$ constant. The remaining terms yield the correction $\delta n_{(m)}$
\begin{equation}\label{eqn:deltan_m}
\partial_{\tau}\left(\frac{\delta n_{(m)}}{\bar{n}_{(m)}}\right)
+\partial_{i}v^{i}
+3\partial_{\tau}C
=0
\end{equation}
which is the relativistic continuity equation.

%\[
%\partial_{\tau}(\frac{\delta n_{(m)}}{\bar{n}_{(m)}})
%=-\frac{1}{2}(2\partial_{\underline{j}}v^{\underline{j}}+\partial_{\tau}h)
%=-\partial_{\underline{j}}\left(
%\frac{\alpha_{(m)}}{4F_{(m)}R}\frac{\dot{a}}{a}e_{j}^{\underline{j}}\Lambda^{j}_{\phantom{-}i}\delta\theta^{i}
%\right)
%-\frac{1}{2}\partial_{\tau}h
%\]\[
%=-\frac{\alpha_{(m)}}{4F_{(m)}R}\frac{\dot{a}}{a}\partial_{\underline{j}}\left(
%e_{j}^{\underline{j}}\Lambda^{j}_{\phantom{-}i}\delta\theta^{i}
%\right)
%-\frac{1}{2}\partial_{\tau}h
%\]

The spin current conservation
\begin{equation}
0
=\nabla_{\mu}\bar{j}_{(s)}^{\mu}
=\partial_{\mu}\bar{j}_{(s)}^{\mu}
+4H\bar{j}_{(s)}^{\tau}.
\end{equation}

The projection of the conservation of the Lie-current $J^{\mu}$, eq~(\ref{eqn:dSDJ=0}), along the $T^{ab}$ Lie elements gives
\[
0
=\nabla_{\mu}\left(j_{(m)}^{\mu}Q_{(m)}^{\alpha\beta}+j_{(s)}^{\mu}Q_{(s)}^{\alpha\beta}\right)
=j_{(m)}^{\mu}\nabla_{\mu}Q_{(m)}^{\alpha\beta}
+j_{(s)}^{\mu}\nabla_{\mu}Q_{(s)}^{\alpha\beta}
\simeq \bar{j}_{(m)}^{\mu}\bar{\nabla}_{\mu}Q_{(m)}^{\alpha\beta}
+\bar{j}_{(s)}^{\mu}\bar{\nabla}_{\mu}Q_{(s)}^{\alpha\beta}
\]
where we made use of eq.~(\ref{eqn:dSDj=0}). The $\underline{i}\tau$ and $\underline{ij}$ components of the above equation are given by
\begin{eqnarray}
\bar{j}_{(m)}^{\mu}\bar{\nabla}_{\mu}Q_{(m)}^{\underline{i}\tau}
+\bar{j}_{(s)}^{\mu}\bar{\nabla}_{\mu}Q_{(s)}^{\underline{i}\tau}
&=&\bar{j}_{(m)}^{\tau}[
\partial_{\tau}Q_{(m)}^{\underline{i}\tau}
+2HQ_{(m)}^{\underline{i}\tau}
]
+H\bar{j}_{(s)}^{\mu}\delta_{\mu\underline{j}}Q_{(s)}^{\underline{ij}}\nonumber\\
&=&\frac{\bar{j}_{(m)}^{\tau}}{a^{2}}\partial_{\tau}(a^{2}Q_{(m)}^{\underline{i}\tau})
+H\bar{j}_{(s)}^{\mu}\delta_{\mu\underline{j}}Q_{(s)}^{\underline{ij}}
=0\label{eqn:DJ^itau}\\
\bar{j}_{(m)}^{\mu}\bar{\nabla}_{\mu}Q_{(m)}^{\underline{ij}}
+\bar{j}_{(s)}^{\mu}\bar{\nabla}_{\mu}Q_{(s)}^{\underline{ij}}
&=&\bar{j}_{(s)}^{\mu}[
\partial_{\mu}Q_{(s)}^{\underline{ij}}
+2\delta^{\tau}_{\mu}HQ_{(s)}^{\underline{ij}}
]
=0\label{eqn:DJ^ij}
\end{eqnarray}
Instead the projection of the conservation of the Lie-current $J^{\mu}$, eq~(\ref{eqn:dSDJ=0}), along the $T^{a}$ Lie elements gives

\begin{eqnarray}
0
&=&\alpha_{(m)}j^{\mu}_{(m)}(\Lambda^{-1}\nabla_{\mu}\Lambda)^{0}_{\phantom{-}b}\nonumber\\
&\simeq& \alpha_{(m)}(\bar{j}^{\mu}_{(m)}+\delta n_{(m)} \bar{u}^{\mu}_{(m)})(\bar{\Lambda}^{-1}\bar{\nabla}_{\mu}\bar{\Lambda})^{0}_{\phantom{-}b}
+\alpha_{(m)}n_{(m)}\delta u^{\mu}_{(m)}(\bar{\Lambda}^{-1}\bar{\nabla}_{\mu}\bar{\Lambda})^{0}_{\phantom{-}b}
+\alpha_{(m)}\bar{j}^{\mu}_{(m)}\delta(\Lambda^{-1}\nabla_{\mu}\Lambda)^{0}_{\phantom{-}b}\nonumber\\
&=&\alpha_{(m)}n_{(m)}\left[
\delta u^{\underline{i}}_{(m)}H\delta^{i}_{\underline{i}}\bar{\Lambda}_{ib}
+\frac{1}{a}\delta(\Lambda^{-1}\nabla_{\tau}\Lambda)^{0}_{\phantom{-}b}
\right]
=\frac{\alpha_{(m)}n_{(m)}}{a}\left[
v^{\underline{i}}_{(m)}H\delta^{i}_{\underline{i}}\bar{\Lambda}_{ib}
+\delta(\Lambda^{-1}\nabla_{\tau}\Lambda)^{0}_{\phantom{-}b}
\right]
\end{eqnarray}

While the above equation is trivially verified at the zero order, at the first order it produces corrections to the group-valued field $\Lambda_{FRW}$, which do not contribute to the first order metric perturbation. Therefore it will not be of any interest in this paper.

Finally from equations~(\ref{eqn:dSDj=0},\ref{eqn:dSj_i},\ref{eqn:dSDJ=0}) it is possible to show \cite{Capasso:2012dr} that the divergence of the energy momentum tensor is indeed zero. This equation is the fluid generalization of the Mathisson-Papapetrou equation for spinning particles.

The last set of equations is given by the linearized Einstein's equation. 

The unperturbed energy-momentum tensor takes the expected form
\[
\bar{T}^{\mu\nu}
=-[\bar{n}_{(m)}-F(\bar{n}_{(m)},\bar{n}_{(s)})]\bar{g}^{\mu\nu}
+\frac{F_{(m)}}{n_{(m)}}j_{(m)}^{\mu}j_{(m)}^{\nu}
=\bar{n}_{(m)}\bar{u}_{(m)}^{\mu}\bar{u}_{(m)}^{\nu}
\]
while its first-order correction is give by
%there is a mistake in \ref{Capasso:2013dr} where a 2 because of the symmetrization is missing
\begin{eqnarray}
\delta T_{\mu\nu}
&=&2\bar{n}_{(m)}\bar{u}_{(m)(\mu}\delta u_{(m)\nu)}
+\delta n_{(m)}\bar{u}_{(m)\mu}\bar{u}_{(m)\nu}
+\alpha_{(s)}q\bar{n}_{(s)}\bar{u}_{(s)\mu}\bar{u}_{(s)\nu}\nonumber\\
&&-4\bar{\nabla}_{\gamma}[\bar{j}_{(m)(\mu}Q_{(m)\nu)}^{\phantom{---}\gamma}]
-4\bar{\nabla}_{\gamma}[\bar{j}_{(s)(\mu}Q_{(s)\nu)}^{\phantom{---}\gamma}]
\end{eqnarray}

\noindent In components we have
\begin{eqnarray}
\delta T_{\tau\tau}
&=&2\bar{n}_{(m)}a^{2}A
+\delta n_{(m)}a^{2}
+\alpha_{(s)}q\bar{n}_{(s)}\left(\bar{u}_{(s)\tau}\right)^{2}
-4\partial_{\underline{i}}\left(\bar{j}_{(m)\tau}\frac{1}{a^{2}}Q_{(m)\underline{i}\tau}\right)\\
\delta T_{\underline{i}\tau}
%=\bar{n}_{(m)}\bar{u}_{(m)\tau}\delta u_{(m)\underline{i}}
%+\alpha_{(s)}q\bar{n}_{(s)}\bar{u}_{(s)\underline{i}}\bar{u}_{(s)\tau}
%-\partial_{\gamma}[\bar{j}_{(m)\tau}Q_{(m)\underline{i}}^{\phantom{---}\gamma}]
%-\frac{\dot{a}}{a}\frac{1}{a^{2}}Q_{(m)\tau\underline{i}}\bar{j}_{(m)\tau}
%+\frac{\dot{a}}{a}Q_{(m)\underline{i}}^{\phantom{---}\tau}\bar{j}_{(m)\tau}
%-\frac{\dot{a}}{a}\bar{j}_{(m)\tau}\frac{1}{a^{2}}Q_{(m)\tau\underline{i}}
%+\frac{\dot{a}}{a}\bar{j}_{(m)\tau}Q_{(m)\underline{i}}^{\phantom{---}\tau}
%-4\frac{\dot{a}}{a}\bar{j}_{(m)\tau}Q_{(m)\underline{i}}^{\phantom{---}\tau}
%-\partial_{\underline{j}}[\bar{j}_{(s)\tau}Q_{(s)\underline{i}}^{\phantom{---}\underline{j}}]
%+\frac{\dot{a}}{a}Q_{(s)\underline{i}}^{\phantom{---}\underline{j}}\bar{j}_{(s)\underline{j}}
&=&\bar{n}_{(m)}\bar{u}_{(m)\tau}\delta u_{(m)\underline{i}}
+\alpha_{(s)}q\bar{n}_{(s)}\bar{u}_{(s)\underline{i}}\bar{u}_{(s)\tau}
-2\partial_{\tau}\left(\bar{j}_{(m)\tau}Q_{(m)\underline{i}}^{\phantom{---}\tau}\right)\nonumber\\
&&-2\partial_{\underline{j}}\left(\bar{j}_{(s)\tau}Q_{(s)\underline{i}}^{\phantom{---}\underline{j}}\right)
+2HQ_{(s)\underline{i}}^{\phantom{---}\underline{j}}\bar{j}_{(s)\underline{j}}\\
\delta T_{\underline{ij}}
&=&\alpha_{(s)}q\bar{n}_{(s)}\bar{u}_{(s)\underline{i}}\bar{u}_{(s)\underline{j}}
-4\partial_{\underline{k}}\left(\bar{j}_{(s)(\underline{i}}Q_{(s)\underline{j})}^{\phantom{---}\underline{k}}\right)
\end{eqnarray}

Using the expression first-order correction to the Einstein tensor found in appendix~\ref{app:curvature} and considering the gauge choice introduced in section~\ref{sec:vorticity}, we find the following linearized Einstein'a equations:

\begin{eqnarray}
\delta G_{\tau\tau}
&=&2\nabla^{2}\Phi
-6H\dot{\Phi}\nonumber\\
&=&8\pi\left[
2\bar{n}_{(m)}a^{2}\Psi
+\delta n_{(m)}a^{2}
+\alpha_{(s)}q\bar{n}_{(s)}\left(\bar{u}_{(s)\tau}\right)^{2}
-4\partial_{\underline{i}}\left(\bar{j}_{(m)\tau}\frac{1}{a^{2}}Q_{(m)\underline{i}\tau}\right)
\right]\\
\delta G_{\underline{i}\tau}
&=&\frac{1}{2}\nabla^{2}\hat{\Phi}_{\underline{i}}
+2\partial_{\underline{i}}\left(H\Psi+\dot{\Phi}\right)\nonumber\\
&=&8\pi\left[
-a^{2}\bar{n}_{(m)}v_{\underline{i}}
+\alpha_{(s)}q\bar{n}_{(s)}\bar{u}_{(s)\underline{i}}\bar{u}_{(s)\tau}
-2\partial_{\tau}\left(\bar{j}_{(m)\tau}Q_{(m)\underline{i}}^{\phantom{---}\tau}\right)
\right.\nonumber\\
&&\left.
-2\partial_{\underline{j}}\left(\bar{j}_{(s)\tau}Q_{(s)\underline{i}}^{\phantom{---}\underline{j}}\right)
+2HQ_{(s)\underline{i}}^{\phantom{---}\underline{j}}\bar{j}_{(s)\underline{j}}
\right]\\
\delta G_{\underline{ij}}
&=&\partial_{\underline{i}}\partial_{\underline{j}}(\Phi-\Psi)
+2H\partial_{(\underline{i}}\hat{\Phi}_{\underline{j})}
+\partial_{(\underline{i}}\dot{\hat{\Phi}}_{\underline{j})}
+\partial^{2}_{\tau}\hat{E}_{\underline{ji}}
-\nabla^{2}\hat{E}_{\underline{ji}}
+2H\partial_{\tau}\hat{E}_{\underline{ij}}
-2\left(2\dot{H}+H^{2}\right)\hat{E}_{\underline{ji}}\nonumber\\
&&-2\left(2\dot{H}+H^{2}\right)\partial_{(\underline{i}}\hat{\Phi}_{\underline{j})}
+\delta_{\underline{ij}}\left[
\nabla^{2}(\Psi-\Phi)
+2\left(2\dot{H}+H^{2}\right)(\Psi+\Phi)
+2\ddot{\Phi}
+H(2\dot{\Psi}+4\dot{\Phi})
\right]\nonumber\\
&=&8\pi\left[
\alpha_{(s)}q\bar{n}_{(s)}\bar{u}_{(s)\underline{i}}\bar{u}_{(s)\underline{j}}
-4\partial_{\underline{k}}\left(\bar{j}_{(s)(\underline{i}}Q_{(s)\underline{j})}^{\phantom{---}\underline{k}}\right)
\right]
\end{eqnarray}

In the following section we shall study the particular case of a stationary spin current.

%%%%%%%%%%%%%%%%%%%%%%%%%%%%%%%%%%%%%%%%%%%
\subsection{Stationary Spin-current}
\label{sec:StationarySpinCurrent}
%%%%%%%%%%%%%%%%%%%%%%%%%%%%%%%%%%%%%%%%%%%

In this section we consider the case in which the spin-current
\[
j_{(s)\mu}
=-\frac{n_{(s)}\alpha_{(s)}}{2F_{(s)}R}\left(
\partial_{\mu}\alpha
-\sin\beta\partial_{\mu}\gamma
\right)
\]
is stationary, that is,
\[
\bar{j}^{\mu}_{(s)}
=\bar{n}_{(s)}a(1,0,0,0).
\]
This implies that the angles $\alpha$, $\beta$ and $\gamma$ are only function of the cosmological time $\tau$ subject to the following condition
\[
\dot{\alpha}-\sin\beta\dot{\gamma}
=-\frac{2F_{(s)}R}{\alpha_{(s)}}a.
\]

The conservation of the current $J^{\mu}$, eq.~(\ref{eqn:DJ^itau},\ref{eqn:DJ^ij}), gives
\[
\bar{j}_{(m)}^{\mu}\bar{\nabla}_{\mu}Q_{(m)}^{\underline{i}\tau}
+\bar{j}_{(s)}^{\mu}\bar{\nabla}_{\mu}Q_{(s)}^{\underline{i}\tau}
=\frac{\bar{j}_{(m)}^{\tau}}{a^{2}}\partial_{\tau}(a^{2}Q_{(m)}^{\underline{i}\tau})
+H\bar{j}_{(s)}^{\mu}\delta_{\mu\underline{j}}Q_{(s)}^{\underline{ij}}
=\frac{\bar{j}_{(m)}^{\tau}}{a^{2}}\partial_{\tau}(a^{2}Q_{(m)}^{\underline{i}\tau})
=0
\]\[
\bar{j}_{(m)}^{\mu}\bar{\nabla}_{\mu}Q_{(m)}^{\underline{ij}}
+\bar{j}_{(s)}^{\mu}\bar{\nabla}_{\mu}Q_{(s)}^{\underline{ij}}
=\bar{j}_{(s)}^{\tau}[
\partial_{\tau}Q_{(s)}^{\underline{ij}}
+2HQ_{(s)}^{\underline{ij}}
]
=0,
\]
That is,
\[
Q_{(m)}^{\underline{i}\tau}
=\frac{K_{(m)}^{\underline{i}\tau}}{a^2}
\qquad
Q_{(s)}^{\underline{ij}}
=\frac{K_{(s)}^{\underline{ij}}}{a^2}
\]
with $K_{(s)}^{\underline{ij}}$ constant, being $\alpha$, $\beta$ and $\gamma$ only function of $\tau$, and $K_{(m)}^{\underline{i}\tau}$ function only of the spatial coordinates (through $\delta\theta^{i}$). This implies that $Q_{(s)}^{ab}$ and then $\mathcal{A}$ in $\Lambda^{}_{FRW}$ are constant quantities. A straightforward solution is give by keeping $\alpha$ as function of only $\tau$ and choosing $\beta=\gamma=0$ which gives
\[
Q_{(s)}^{ab}
=\frac{\alpha_{(s)}}{2R}\left(\begin{array}{cccc}
0&0&0&0\\
0&0&0&0 \\
0&0&0&1\\
0&0&-1&0
\end{array}\right)
\]
Finally from the linearized Einstein's equations we have
\begin{equation}
2\nabla^{2}\Phi
-6H\dot{\Phi}
=8\pi\left[
2\bar{n}_{(m)}a^{2}\Psi
+\delta n_{(m)}a^{2}
+\alpha_{(s)}q\bar{n}_{(s)}a^{2}
-4\partial_{\underline{i}}\left(\bar{j}_{(m)\tau}\frac{1}{a^{2}}Q_{(m)\underline{i}\tau}\right)
\right],
\end{equation}
\begin{equation}
\frac{1}{2}\nabla^{2}\hat{\Phi}_{\underline{i}}
+2\partial_{\underline{i}}\left(H\Psi+\dot{\Phi}\right)
=8\pi\left[
-a^{2}\bar{n}_{(m)}v_{\underline{i}}
-2\partial_{\tau}\left(\bar{j}_{(m)\tau}Q_{(m)\underline{i}}^{\phantom{---}\tau}\right)
\right],
\end{equation}
and
\begin{eqnarray}
&\partial_{\underline{i}}\partial_{\underline{j}}(\Phi-\Psi)
+2H\partial_{(\underline{i}}\hat{\Phi}_{\underline{j})}
+\partial_{(\underline{i}}\dot{\hat{\Phi}}_{\underline{j})}
+\partial^{2}_{\tau}\hat{E}_{\underline{ji}}
-\nabla^{2}\hat{E}_{\underline{ji}}
+2H\partial_{\tau}\hat{E}_{\underline{ij}}
-2\left(2\dot{H}+H^{2}\right)\hat{E}_{\underline{ji}}
-2\left(2\dot{H}+H^{2}\right)\partial_{(\underline{i}}\hat{E}_{\underline{j})}&\nonumber\\
&+\delta_{\underline{ij}}\left[
\nabla^{2}(\Psi-\Phi)
+2\left(2\dot{H}+H^{2}\right)(\Psi+\Phi)
+2\ddot{\Phi}
+H(2\dot{\Psi}+4\dot{\Phi})
\right]
=0&
\end{eqnarray}

Expanding the $\underline{i}\tau$ component
\[
\frac{1}{2}\nabla^{2}\hat{\Phi}_{\underline{i}}
+2\partial_{\underline{i}}\left(H\Psi+\dot{\Phi}\right)
=8\pi\left[
-a^{2}\bar{n}_{(m)}v_{\underline{i}}
-2\partial_{\tau}\left(\bar{j}_{(m)\tau}Q_{(m)\underline{i}}^{\phantom{---}\tau}\right)
\right]
\]\[
=8\pi\left[
-a^{2}\bar{n}_{(m)}v_{\underline{i}}
+4H\left(\frac{\bar{j}_{(m)\tau}}{a^{2}}Q_{(m)\underline{i}\tau}\right)
\right]
=8\pi\left[
-a^{2}\bar{n}_{(m)}v_{\underline{i}}
+2\bar{n}_{(m)}\left(
\frac{2H}{a}e_{\underline{i}}^{i}e_{\tau}^{0}\frac{\alpha_{(m)}}{8R}\Lambda_{ij}\delta\theta^{j}
\right)
\right]
\]\[
=8\pi\left[
-a^{2}\bar{n}_{(m)}v_{\underline{i}}
+2\bar{n}_{(m)}a^{2}\left(
v_{\underline{i}}
+\frac{\alpha_{(m)}}{4R}\frac{\partial_{\underline{i}}\theta_{0}}{a}
\right)
\right]
=8\pi\left[
a^{2}\bar{n}_{(m)}v_{\underline{i}}
+2\bar{n}_{(m)}a^{2}\frac{\alpha_{(m)}}{4R}\frac{\partial_{\underline{i}}\theta_{0}}{a}
\right],
\]
taking the spatial derivative and anti-symmetrizing with respect to the two spatial indexes, we find an equation for the vectorial perturbation:
\[
\frac{1}{2}\nabla^{2}\partial_{[j}\hat{\Phi}_{\underline{i}]}
=8\pi a^{2}\bar{n}_{(m)}\partial_{[j}v_{\underline{i}]}
\]
The field $\hat{\Phi}$ is divergenceless therefore the above equations shows how it is completely determined by the vorticity of the fluid.

%%%%%%%%%%%%%%%%%%%%%%%%%%%%%%%%%%%%%%%%%%%
\section{Conclusion}
\label{sec:conclusion}
%%%%%%%%%%%%%%%%%%%%%%%%%%%%%%%%%%%%%%%%%%%

In this paper we used the model introduced in \cite{Capasso:2012dr} to show how it is possible to study the presence of vorticity and spin-density during the cosmological evolution of the universe. We restricted our consideration to the only presence of a matte fluid with spin. The vorticity as well as the spin-density of the fluid are introduced as a first order perturbation of an ideal fluid.

While a more general study requires numerical simulations, in this paper we considered the simpler case of a stationary spin-current. From this assumption it has been shown how the spin-density does not contribute to any cosmological perturbation, while the vorticity of the fluid corresponds to a vectorial perturbation.

From a more general analysis it is expected, as the equations of motion suggest, the spin-density to be a possible source of tensorial perturbations.

%%%%%%%%%%%%%%%%%%%%%%%%%%%%%%%%%%%%%
%\begin{center}
%\textbf{Acknowledgements}
%\end{center}
%%%%%%%%%%%%%%%%%%%%%%%%%%%%%%%%%%%%%

%We are indebted to V.P. Nair for some very useful discussions throughout this project. The work of D.S. is supported in part by US National Science Foundation grant PHY-0855582 and Lehman College CUNY Science Fellowship.

\vspace{2cm}
%%%%%%%%%%%%%%%%
\appendix
%%%%%%%%%%%%%%%%
\begin{center}
\large{\textbf{Appendix}}
\end{center}

%%%%%%%%%%%%%%%%%%%%%%%%%%%%%%%%%%%%%%%%%%%%%%%%%%%%%
\section{Christoffel Symbols, Curvature, et cetera}
\label{app:curvature}
%%%%%%%%%%%%%%%%%%%%%%%%%%%%%%%%%%%%%%%%%%%%%%%%%%%%%

In this appendix we shall derive the Christoffel symbols, the Riemannian curvature, the Ricci tensor, the Ricci scalar, and the Einstein tensor for the metric (\ref{eqn:flatmetric}), as well as their first order correction from the perturbed metric (\ref{eqn:perturbedmetric})
\[
ds^{2}
=a^{2}(\tau)[(1+2A)d\tau^{2}-2B_{i}dx^{i}d\tau-(\delta_{ij}+h_{ij})dx^{i}dx^{j}].
\]

The Christoffel symbols are straightforward
\[
\Gamma^{\alpha}_{\mu\beta}
%=\frac{1}{a}(\partial_{\mu}a\delta^{\alpha}_{\beta}+\partial_{\beta}a\delta^{\alpha}_{\mu}-\eta_{\mu\beta}\eta^{\alpha\gamma}\partial_{\gamma}a)
=\delta^{\alpha}_{\beta}\partial_{\mu}\ln a
+\delta^{\alpha}_{\mu}\partial_{\beta}\ln a
-\eta_{\mu\beta}\eta^{\alpha\gamma}\partial_{\gamma}\ln a
\]
giving the following non-zero Christoffel symbols
\[
\bar{\Gamma}^{\alpha}_{\tau\beta}=\bar{\Gamma}^{\alpha}_{\beta\tau}=\frac{\dot{a}}{a}\delta^{\alpha}_{\beta}
\qquad
\bar{\Gamma}^{\tau}_{ij}=\frac{\dot{a}}{a}\delta_{ij}
\]
The Riemannian tensor will then be given by
\[
R^{\alpha}_{\phantom{-}\beta\mu\nu}
%=2\partial_{[\mu}\Gamma^{\alpha}_{\nu]\beta}+2\Gamma^{\alpha}_{[\mu|\gamma|}\Gamma^{\gamma}_{\nu]\beta}
%=2\partial_{[\mu}[\delta^{\alpha}_{|\beta|}\partial_{\nu]}\ln a
%+\delta^{\alpha}_{\nu]}\partial_{\beta}\ln a
%-\eta_{\nu]\beta}\eta^{\alpha\gamma}\partial_{\gamma}\ln a]
%\]\[
%+2[
%\delta^{\alpha}_{\gamma}\partial_{[\mu}\ln a
%+\delta^{\alpha}_{[\mu}\partial_{|\gamma|}\ln a
%-\eta_{[\mu|\gamma|}\eta^{\alpha\delta}\partial_{|\delta|}\ln a
%][
%\delta^{\gamma}_{|\beta|}\partial_{\nu]}\ln a
%+\delta^{\gamma}_{\nu]}\partial_{\beta}\ln a
%-\eta_{\nu]\beta}\eta^{\gamma\rho}\partial_{\rho}\ln a
%]
%\]\[
%=2[
%\delta^{\alpha}_{[\nu}\partial_{\mu]}\partial_{\beta}\ln a
%-\eta_{\beta[\nu}\eta^{\alpha\gamma}\partial_{\mu]}\partial_{\gamma}\ln a]
%\]\[
%+2[
%\delta^{\alpha}_{[\mu}\delta^{\gamma}_{\nu]}\partial_{\gamma}\ln a\partial_{\beta}\ln a
%-\delta^{\alpha}_{[\mu}\eta_{\nu]\beta}\eta^{\gamma\rho}\partial_{\gamma}\ln a\partial_{\rho}\ln a
%+\eta^{\alpha\delta}\eta_{\beta[\nu}\partial_{\mu]}\ln a\partial_{\delta}\ln a
%]
%\]\[
%=2[
%\delta^{\tau}_{\beta}\delta^{\alpha}_{[\nu}\delta^{\tau}_{\mu]}\dot{H}
%-\eta^{\alpha\tau}\eta_{\beta[\nu}\delta^{\tau}_{\mu]}\dot{H}]
%+2[
%\delta^{\alpha}_{[\mu}\delta^{\tau}_{\nu]}\delta^{\tau}_{\beta}H^{2}
%-\delta^{\alpha}_{[\mu}\eta_{\nu]\beta}H^{2}
%+\eta^{\alpha\tau}\eta_{\beta[\nu}\delta^{\tau}_{\mu]}H^{2}
%]
%\]\[
=2[
\delta^{\tau}_{\beta}\delta^{\alpha}_{[\nu}\delta^{\tau}_{\mu]}
-\eta^{\alpha\tau}\eta_{\beta[\nu}\delta^{\tau}_{\mu]}
](\dot{H}-H^{2})
-2\delta^{\alpha}_{[\mu}\eta_{\nu]\beta}H^{2}
\]
while the Ricci tensor is
\[
R_{\beta\nu}
%=R^{\alpha}_{\phantom{-}\beta\alpha\nu}
%\]\[
%=2[
%\delta^{\alpha}_{[\nu}\partial_{\alpha]}\partial_{\beta}\ln a
%-\eta_{\beta[\nu}\eta^{\alpha\gamma}\partial_{\alpha]}\partial_{\gamma}\ln a]
%\]\[
%+2[
%\delta^{\alpha}_{[\alpha}\delta^{\gamma}_{\nu]}\partial_{\gamma}\ln a\partial_{\beta}\ln a
%-\delta^{\alpha}_{[\alpha}\eta_{\nu]\beta}\eta^{\gamma\rho}\partial_{\gamma}\ln a\partial_{\rho}\ln a
%+\eta^{\alpha\delta}\eta_{\beta[\nu}\partial_{\alpha]}\ln a\partial_{\delta}\ln a
%]
%\]\[
%=-2\partial_{\nu}\partial_{\beta}\ln a
%-\eta_{\beta\nu}\eta^{\alpha\gamma}\partial_{\alpha}\partial_{\gamma}\ln a
%+2\partial_{\nu}\ln a\partial_{\beta}\ln a
%-2\eta_{\nu\beta}\eta^{\gamma\rho}\partial_{\gamma}\ln a\partial_{\rho}\ln a
%\]\[
%=-2\delta^{\tau}_{\nu}\delta^{\tau}_{\beta}\left[
%\frac{\ddot{a}}{a}-2\frac{\dot{a}^{2}}{a^{2}}
%\right]
%-\eta_{\nu\beta}[
%\frac{\ddot{a}}{a}+\frac{\dot{a}^{2}}{a^{2}}
%]
=-2\delta^{\tau}_{\nu}\delta^{\tau}_{\beta}(\dot{H}-H^{2})
-\eta_{\nu\beta}(\dot{H}+2H^{2})
\]
and the Ricci scalar is
\[
R
%=g^{\beta\nu}R_{\beta\nu}
%=\frac{1}{a^{2}}\eta^{\beta\nu}R_{\beta\nu}
%=-\frac{6}{a^{2}}\left[
%\eta^{\beta\nu}\partial_{\nu}\partial_{\beta}\ln a
%+\eta^{\gamma\rho}\partial_{\gamma}\ln a\partial_{\rho}\ln a
%\right]
=-\frac{6}{a^{2}}\frac{\ddot{a}}{a}
\]
Finally we can write the Einstein tensor
\[
G_{\alpha\beta}
=R_{\alpha\beta}-\frac{1}{2}g_{\alpha\beta}R
%=R_{\alpha\beta}-\frac{a^{2}}{2}\eta_{\alpha\beta}R
%\]\[
=\eta_{\alpha\beta}\left[
2\frac{\ddot{a}}{a}-\frac{\dot{a}^{2}}{a^{2}}
\right]
-2\delta^{\tau}_{\alpha}\delta^{\tau}_{\beta}\left[
\frac{\ddot{a}}{a}-2\frac{\dot{a}^{2}}{a^{2}}
\right]
\]
which has the following non-zero components
\[
G_{\tau\tau}
=3\frac{\dot{a}^{2}}{a^{2}}
\qquad
G_{ij}
=\left[
2\frac{\ddot{a}}{a}-\frac{\dot{a}^{2}}{a^{2}}
\right]\eta_{ij}.
\]

The first order correction to the Christoffel symbols
\[
\delta\Gamma^{\alpha}_{\mu\beta}
=\frac{1}{2a^{2}}\eta^{\alpha\gamma}[
\bar{\nabla}_{\mu}\delta g_{\gamma\beta}
+\bar{\nabla}_{\beta}\delta g_{\mu\gamma}
-\bar{\nabla}_{\gamma}\delta g_{\mu\beta}
]
%\]\[
%=\frac{1}{2a^{2}}\eta^{\alpha\gamma}[
%\partial_{\mu}\delta g_{\gamma\beta}
%-\bar{\Gamma}^{\delta}_{\mu\gamma}\delta g_{\delta\beta}
%-\bar{\Gamma}^{\delta}_{\mu\beta}\delta g_{\gamma\delta}
%+\partial_{\beta}\delta g_{\mu\gamma}
%-\bar{\Gamma}^{\delta}_{\beta\mu}\delta g_{\delta\gamma}
%-\bar{\Gamma}^{\delta}_{\beta\gamma}\delta g_{\mu\delta}
%-\partial_{\gamma}\delta g_{\mu\beta}
%+\bar{\Gamma}^{\delta}_{\gamma\mu}\delta g_{\delta\beta}
%+\bar{\Gamma}^{\delta}_{\gamma\beta}\delta g_{\mu\delta}
%]
%\]\[
%=\frac{1}{2a^{2}}\eta^{\alpha\gamma}[
%\partial_{\mu}\delta g_{\gamma\beta}
%+\partial_{\beta}\delta g_{\mu\gamma}
%-\partial_{\gamma}\delta g_{\mu\beta}
%-2\bar{\Gamma}^{\delta}_{\mu\beta}\delta g_{\gamma\delta}
%]
%\]\[
%=\frac{1}{2a^{2}}\eta^{\alpha\gamma}[
%2\frac{\delta g_{\gamma\beta}}{a}\dot{a}\delta_{\mu}^{\tau}
%+2\frac{\delta g_{\mu\gamma}}{a}\dot{a}\delta_{\beta}^{\tau}
%-2\frac{\delta g_{\mu\beta}}{a}\dot{a}\delta_{\gamma}^{\tau}
%+a^{2}\partial_{\mu}\frac{\delta g_{\gamma\beta}}{a^{2}}
%+a^{2}\partial_{\beta}\frac{\delta g_{\mu\gamma}}{a^{2}}
%-a^{2}\partial_{\gamma}\frac{\delta g_{\mu\beta}}{a^{2}}
%-2\frac{\dot{a}}{a}\delta_{\mu}^{\tau}\delta g_{\gamma\beta}
%-2\frac{\dot{a}}{a}\delta_{\beta}^{\tau}\delta_{\mu}^{i}\delta g_{\gamma i}
%-2\frac{\dot{a}}{a}\delta_{ij}\delta_{\mu}^{i}\delta_{\beta}^{j}\delta g_{\gamma\tau}
%]
\]\[
=\frac{1}{2a^{2}}\eta^{\alpha\gamma}[
-2\frac{\delta g_{\mu\beta}}{a}\dot{a}\delta_{\gamma}^{\tau}
+2\frac{\delta g_{\tau\gamma}}{a}\dot{a}\delta_{\beta}^{\tau}\delta_{\mu}^{\tau}
+a^{2}\partial_{\mu}\frac{\delta g_{\gamma\beta}}{a^{2}}
+a^{2}\partial_{\beta}\frac{\delta g_{\mu\gamma}}{a^{2}}
-a^{2}\partial_{\gamma}\frac{\delta g_{\mu\beta}}{a^{2}}
-2\frac{\dot{a}}{a}\delta_{ij}\delta_{\mu}^{i}\delta_{\beta}^{j}\delta g_{\gamma\tau}
]
\]
has the following non-zero terms
\[
\delta\Gamma^{\tau}_{ij}
%=\frac{1}{2a^{2}}[
%-2\frac{\delta g_{ij}}{a}\dot{a}
%+a^{2}\partial_{i}\frac{\delta g_{\tau j}}{a^{2}}
%+a^{2}\partial_{j}\frac{\delta g_{i\tau}}{a^{2}}
%-a^{2}\partial_{\tau}\frac{\delta g_{ij}}{a^{2}}
%-2\frac{\dot{a}}{a}\delta_{ij}\delta g_{\tau\tau}
%]
%\]\[
=\frac{\dot{a}}{a}h_{ij}
-\partial_{(i}B_{j)}
+\frac{1}{2}\partial_{\tau}h_{ij}
-2H\delta_{ij}A
\]

\[
\delta\Gamma^{\tau}_{\tau\tau}
=\partial_{\tau}A
\qquad
\delta\Gamma^{\tau}_{\tau j}
=HB_{j}
+\partial_{j}A
\]

%\[
%\delta\Gamma^{k}_{\mu\beta}
%=-\frac{1}{2a^{2}}[
%2\frac{\delta g_{\tau k}}{a}\dot{a}\delta_{\beta}^{\tau}\delta_{\mu}^{\tau}
%+a^{2}\partial_{\mu}\frac{\delta g_{k\beta}}{a^{2}}
%+a^{2}\partial_{\beta}\frac{\delta g_{\mu k}}{a^{2}}
%-a^{2}\partial_{k}\frac{\delta g_{\mu\beta}}{a^{2}}
%+2\frac{\dot{a}}{a}\delta_{ij}\delta_{\mu}^{i}\delta_{\beta}^{j}a^{2}B_{k}
%]
%\]
%
%\[
%\delta\Gamma^{k}_{\tau\beta}
%=-\frac{1}{2a^{2}}[
%2\frac{\delta g_{\tau k}}{a}\dot{a}\delta_{\beta}^{\tau}
%+a^{2}\partial_{\tau}\frac{\delta g_{k\beta}}{a^{2}}
%+a^{2}\partial_{\beta}\frac{\delta g_{\tau k}}{a^{2}}
%-a^{2}\partial_{k}\frac{\delta g_{\tau\beta}}{a^{2}}
%]
%\]\[
%=-\frac{1}{2a^{2}}[
%-2a^{2}\frac{B_{k}}{a}\dot{a}\delta_{\beta}^{\tau}
%+a^{2}\partial_{\tau}\frac{\delta g_{k\beta}}{a^{2}}
%-a^{2}\partial_{\beta}B_{k}
%-a^{2}\partial_{k}\frac{\delta g_{\tau\beta}}{a^{2}}
%]
%\]

\[
\delta\Gamma^{k}_{\tau\tau}
=HB_{k}
+\partial_{\tau}B_{k}
+\partial_{k}A
\qquad
\delta\Gamma^{k}_{\tau j}
=\frac{1}{2}\partial_{\tau}h_{kj}
+\partial_{[j}B_{k]}
\]

\[
\delta\Gamma^{k}_{ij}
=\frac{1}{2}[
\partial_{i}h_{kj}
+\partial_{j}h_{ik}
-\partial_{k}h_{ij}
-2H\delta_{ij}B_{k}
]
\]

%The first order correction to the Riemannian tensor
%\[
%\delta R^{\alpha}_{\phantom{-}\beta\mu\nu}
%=2\partial_{[\mu}\delta\Gamma^{\alpha}_{\nu]\beta}
%+\delta\Gamma^{\alpha}_{\mu\gamma}\bar{\Gamma}^{\gamma}_{\nu\beta}
%+\bar{\Gamma}^{\alpha}_{\mu\gamma}\delta\Gamma^{\gamma}_{\nu\beta}
%-\delta\Gamma^{\alpha}_{\nu\gamma}\bar{\Gamma}^{\gamma}_{\mu\beta}
%-\bar{\Gamma}^{\alpha}_{\nu\gamma}\delta\Gamma^{\gamma}_{\mu\beta}
%\]
%which gives
%\[
%\delta R^{\alpha}_{\phantom{-}\beta\mu\nu}
%=2\bar{\nabla}_{[\mu}\delta\Gamma^{\alpha}_{\nu]\beta}
%\]

The first order correction to the Ricci tensor can be explicitly derived as follow

\[
\delta R_{\beta\nu}
=\delta R^{\alpha}_{\phantom{-}\beta\alpha\nu}
=2\bar{\nabla}_{[\alpha}\delta\Gamma^{\alpha}_{\nu]\beta}
=\bar{g}^{\alpha\gamma}\bar{\nabla}_{[\alpha}\left[
\bar{\nabla}_{\nu]}\delta g_{\gamma\beta}
+\bar{\nabla}_{|\beta|}\delta g_{\nu]\gamma}
-\bar{\nabla}_{|\gamma|}\delta g_{\nu]\beta}
\right]
\]\[
=\bar{g}^{\alpha\gamma}\bar{\nabla}_{\alpha}\bar{\nabla}_{(\nu}\delta g_{\beta)\gamma}
-\frac{1}{2}\bar{g}^{\alpha\gamma}\bar{\nabla}_{\alpha}\bar{\nabla}_{\gamma}\delta g_{\nu\beta}
-\frac{1}{2}\bar{\nabla}_{\nu}\bar{\nabla}_{\beta}\delta g
\]

\noindent which gives the following non zero-components
\[
\delta R_{\tau\tau}
=3H(\dot{\Psi}+\dot{\Phi})
+\nabla^{2}\Psi
+3\ddot{\Phi}
+3\ddot{H}(B-\dot{E})+6\dot{H}(\dot{B}-\ddot{E})
\]

\[
\delta R_{\underline{i}\tau}
=\frac{1}{2}\nabla^{2}\hat{\Phi}_{\underline{i}}
+2H\partial_{\underline{i}}\Psi
+2\partial_{\underline{i}}\dot{\Phi}
+(\dot{H}+2H^{2})(\hat{B}_{\underline{i}}+\partial_{\underline{i}}\dot{E})
+3\dot{H}\partial_{\underline{i}}(B-\dot{E})
\]

\[
\delta R_{\underline{ij}}
=\bar{g}^{\alpha\gamma}\bar{\nabla}_{\alpha}\bar{\nabla}_{(\underline{j}}\delta g_{\underline{i})\gamma}
-\frac{1}{2}\bar{g}^{\alpha\gamma}\bar{\nabla}_{\alpha}\bar{\nabla}_{\gamma}\delta g_{\underline{ji}}
-\frac{1}{2}\bar{\nabla}_{\underline{j}}\bar{\nabla}_{\underline{i}}\delta g
\]\[
=\partial_{\underline{i}}\partial_{\underline{j}}(\Phi-\Psi)
+\left(
-\ddot{\Phi}
+\nabla^{2}\Phi
-2\left(\dot{H}+2H^{2}\right)\left(\Phi+\Psi\right)
-H\dot{\Psi}
-5H\dot{\Phi}
\right)\delta_{\underline{ij}}
\]\[
+2H\partial_{(\underline{i}}[\dot{\hat{E}}_{\underline{j})}-\hat{B}_{\underline{j})}]
+\partial_{(\underline{i}}[\ddot{\hat{E}}_{\underline{j})}-\dot{\hat{B}}_{\underline{j})}]
+\partial^{2}_{\tau}\hat{E}_{\underline{ji}}
-\nabla^{2}\hat{E}_{\underline{ji}}
+2H\partial_{\tau}\hat{E}_{\underline{ij}}
+2\left(\dot{H}+2H^{2}\right)\hat{E}_{\underline{ji}}
\]\[
-(4H\dot{H}+\ddot{H})(B-\dot{E})\delta_{\underline{ji}}
+2\left(\dot{H}+2H^{2}\right)\left(
\partial_{\underline{i}}\partial_{\underline{j}}E
+\partial_{(\underline{i}}\hat{E}_{\underline{j})}
\right)
\]

\noindent First order correction to the Ricci scalar is given by the following expression
\[
\delta R
=\delta g^{\beta\nu}\bar{R}_{\beta\nu}
+\bar{g}^{\beta\nu}\delta R_{\beta\nu}
\]\[
%=\frac{6}{a^{2}}\left[
%A\dot{H}
%+C(\dot{H}+2H^{2})
%\right]
%+\frac{1}{a^{2}}\left[
%3H(\dot{\Psi}+\dot{\Phi})
%+\nabla^{2}\Psi
%+3\ddot{\Phi}
%\textcolor{red}{+3\ddot{H}(B-\dot{E})+6\dot{H}(\dot{B}-\ddot{E})}
%\right]
%\]\[
%-\frac{1}{a^{2}}\left[
%\nabla^{2}(\Phi-\Psi)
%+3\left(
%-\ddot{\Phi}
%+\nabla^{2}\Phi
%-2\left(\dot{H}+2H^{2}\right)\left(\Phi+\Psi\right)
%-H\dot{\Psi}
%-5H\dot{\Phi}
%\right)
%-3(4H\dot{H}+\ddot{H})(B-\dot{E})
%+2\left(\dot{H}+2H^{2}\right)\nabla^{2}E
%\right]
%\]\[
%=\frac{6}{a^{2}}\left[
%\dot{H}\Psi
%-(\dot{H}+2H^{2})\Phi
%+(\ddot{H}-2H^{3})(B-\dot{E})
%\right]
%\]\[
%+\frac{1}{a^{2}}\left[
%3H(\dot{\Psi}+\dot{\Phi})
%+\nabla^{2}\Psi
%+3\ddot{\Phi}
%\right]
%\]\[
%-\frac{1}{a^{2}}\left[
%\nabla^{2}(\Phi-\Psi)
%+3\left(
%-\ddot{\Phi}
%+\nabla^{2}\Phi
%-2\left(\dot{H}+2H^{2}\right)\left(\Phi+\Psi\right)
%-H\dot{\Psi}
%-5H\dot{\Phi}
%\right)
%\right]
%\]\[
=\frac{1}{a^{2}}\left[
2\nabla^{2}\Psi
-4\nabla^{2}\Phi
+12\left(\dot{H}+H^{2}\right)\Psi
+6\ddot{\Phi}
+6H(\dot{\Psi}+3\dot{\Phi})
\right]
+\frac{6}{a^{2}}\left[
(\ddot{H}-2H^{3})(B-\dot{E})
\right]
\]

And finally the first order correction to the Einstein tensor
\[
\delta G_{\beta\nu}
=\delta R_{\beta\nu}
-\frac{a^{2}}{2}\eta_{\beta\nu}\delta R
-\frac{1}{2}\delta g_{\beta\nu}R
\]
gives the following non-zero components
\[
\delta G_{\tau\tau}
%=\delta R_{\tau\tau}
%-\frac{a^{2}}{2}\delta R
%-\frac{1}{2}\delta g_{\tau\tau}R
%\]\[
%=3H(\dot{\Psi}+\dot{\Phi})
%+\nabla^{2}\Psi
%+3\ddot{\Phi}
%+3\ddot{H}(B-\dot{E})+6\dot{H}(\dot{B}-\ddot{E})
%\]\[
%-\left[
%\nabla^{2}\Psi
%-2\nabla^{2}\Phi
%+6\left(\dot{H}+H^{2}\right)\Psi
%+3\ddot{\Phi}
%+3H(\dot{\Psi}+3\dot{\Phi})
%\right]
%-3\left[
%(\ddot{H}-2H^{3})(B-\dot{E})
%\right]
%+6(\dot{H}+H^{2})A
%\]\[
%=-\left[
%-2\nabla^{2}\Phi
%+6\left(\dot{H}+H^{2}\right)\Psi
%+6H\dot{\Phi}
%\right]
%+6(\dot{H}+H^{2})[A+(\dot{B}-\ddot{E})+H(B-\dot{E})]
%-6H^{2}(\dot{B}-\ddot{E})
%-6H^{2}\dot{H}(B-\dot{E})
%\]\[
=2\nabla^{2}\Phi
-6H\dot{\Phi}
-6H^{2}(\dot{B}-\ddot{E})
-6H^{2}\dot{H}(B-\dot{E})
\]

\[
\delta G_{\underline{i}\tau}
%=\delta R_{\underline{i}\tau}
%+\frac{a^{2}}{2}B_{\underline{i}}R
%\]\[
%=\frac{1}{2}\nabla^{2}\hat{\Phi}_{\underline{i}}
%+2H\partial_{\underline{i}}\Psi
%+2\partial_{\underline{i}}\dot{\Phi}
%\textcolor{red}{+(\dot{H}+2H^{2})(\hat{B}_{\underline{i}}+\partial_{\underline{i}}\dot{E})
%+3\dot{H}\partial_{\underline{i}}(B-\dot{E})}
%-3(\dot{H}+H^{2})\partial_{\underline{i}}B
%-3(\dot{H}+H^{2})\hat{B}_{\underline{i}}
%\]\[
=\frac{1}{2}\nabla^{2}\hat{\Phi}_{\underline{i}}
+2H\partial_{\underline{i}}\Psi
+2\partial_{\underline{i}}\dot{\Phi}
-(2\dot{H}+H^{2})\hat{B}_{\underline{i}}
-2(\dot{H}-H^{2})\partial_{\underline{i}}\dot{E}
-3H^{2}\partial_{\underline{i}}B
\]

\[
\delta G_{\underline{ij}}
=\delta R_{\underline{ij}}
+\frac{a^{2}}{2}\delta_{\underline{ij}}\delta R
+\frac{a^{2}}{2}h_{\underline{ij}}\bar{R}
\]\[
=\partial_{\underline{i}}\partial_{\underline{j}}(\Phi-\Psi)
+2H\partial_{(\underline{i}}\hat{\Phi}_{\underline{j})}
+\partial_{(\underline{i}}\dot{\hat{\Phi}}_{\underline{j})}
+\partial^{2}_{\tau}\hat{E}_{\underline{ji}}
-\nabla^{2}\hat{E}_{\underline{ji}}
+2H\partial_{\tau}\hat{E}_{\underline{ij}}
-2\left(2\dot{H}+H^{2}\right)\hat{E}_{\underline{ji}}
\]\[
-2\left(2\dot{H}+H^{2}\right)\left(
\partial_{\underline{i}}\partial_{\underline{j}}E
+\partial_{(\underline{i}}\hat{E}_{\underline{j})}
\right)
\]\[
+\delta_{\underline{ij}}\left[
\nabla^{2}(\Psi-\Phi)
+2\left(2\dot{H}+H^{2}\right)(\Psi+\Phi)
+2\ddot{\Phi}
+H(2\dot{\Psi}+4\dot{\Phi})
+(2\ddot{H}+2H\dot{H})(B-\dot{E})
\right]
\]

In the above expressions no gauge choice was made. Each component has also been written using the Bardeen's variables (\ref{eqn:Bardeen}). Notice that the remaining terms in each expression not expressed in terms of the Bardeen's variable reproduce the correct transformation properties of each term.

%%%%%%%%%%%%%%%%%%%%%%%%%%%%%%%%%%%%%%%%%%%%%%%%%%%%%
\section{deSitter Group Generators}
\label{app:dSgenerators}
%%%%%%%%%%%%%%%%%%%%%%%%%%%%%%%%%%%%%%%%%%%%%%%%%%%%%

In this appendix we list the convention used in this article for the representation of the Lia algebra associated with the de Sitter group.

The generators of $so(4,1)$ are constructed, similarly to $so(3,1)$, starting the four-dimensional Dirac's gamma matrices
\[
\{\gamma^{a},\gamma^{b}\}=2\eta^{ab}
\]
with the addition of $\gamma^{5}$. Such algebra can be rewritten in a shortened notation as
\[
\{\gamma^{A},\gamma^{B}\}=2\eta^{AB}
\]
where $A$ takes values $\{0,1,2,3,5\}$ and $\eta^{AB}=\{1,-1,-1,-1,-1\}$.

Similarly to the case of the Lorentz group the de Sitter group is generated from Lie algebra generator defined as follows
\begin{eqnarray*}
T_{AB}=\frac{i}{8}\left[\gamma_A,\gamma_B\right]\qquad\textrm{with}\qquad T_{A5}=-T_{5A}\equiv T_{A}
\end{eqnarray*}

The commutation relations between the Lie Generators are
\[
2[T_{AB},T_{CD}]
=i\eta_{BC}T_{AD}-i\eta_{AC}T_{BD}-i\eta_{BD}T_{AC}+i\eta_{AD}T_{BC},
\]
that is,
\[
2[T_{ab},T_{cd}]
=i\eta_{bc}T_{ad}
-i\eta_{ac}T_{bd}
-i\eta_{bd}T_{ac}
+i\eta_{ad}T_{bc}
\]\[
2[T_{ab},T_{c}]
=i\eta_{bc}T_{a}
-i\eta_{ac}T_{b} \qquad
2[T_{a},T_{c}]
=iT_{ac}
\]

The normalized traces of the Lie generators are given by
\[
2\tr[T_{AB}T_{CD}]
=\frac{1}{2}(\eta_{AC}\eta_{BD}-\eta_{AD}\eta_{BC}).
\]

\end{document}